\documentclass[12pt]{article}

\def\fun#1#2{\lower3.6pt\vbox{\baselineskip0pt\lineskip.9pt
\ialign{$\mathsurround=0pt#1\hfil##\hfil$\crcr#2\crcr\sim\crcr}}}
\usepackage{amssymb}
\usepackage{amsmath}
\usepackage{graphicx}
\usepackage{amscd,multicol}

\title{Mass of the higgs versus fourth generation masses}
\author{V.A. Novikov\thanks{email: novikov@heron.itep.ru},
L.B. Okun\thanks{email: okun@heron.itep.ru}, \\ ITEP, Moscow,
Russia
\\
 A.N. Rozanov\thanks{email: rozanov@cppm.in2p3.fr}, \\
CPPM, IN2P3, CNRS, Univ. Mediteranee, Marseilles, France \\
 and
ITEP, Moscow, Russia
\\
  M.I. Vysotsky\thanks{email: vysotsky@heron.itep.ru} \\
ITEP, Moscow, Russia  }
\date{}
\begin{document}
\maketitle

\begin{abstract}

The predicted value of the higgs mass $m_H$ is analyzed  assuming
the existence of the fourth generation of leptons ($N, E$) and
quarks ($U, D$).
 The steep and flat directions are found in the
five-dimensional parameter space: $m_H$, $m_U$, $m_D$, $m_N$,
$m_E$. The LEPTOP fit of the precision electroweak data is
compatible (in particular) with $m_H \sim 300$ GeV, $m_N \sim 50$
GeV, $m_E \sim 100$ GeV, $m_U +m_D \sim 500$ GeV, and $|m_U -m_D|
\sim 75$ GeV. The quality of fits drastically improves when the data
on b- and c-quark asymmetries and new NuTeV data on deep inelastic
scattering are ignored.
\end{abstract}

It is well known that in the framework of Standard Model the fit
of electroweak precision data results in prediction of light
higgs, the central value of its mass being lower than the direct
lower limit set by LEP II \cite{1}. One possible way to raise the
predicted value of $m_H$ is to assume the existence of fourth
generation of leptons and quarks, \cite{4, 3} . Implications of
extra quark-lepton generations for precision data were studied in
a number of papers \cite{4} - \cite{999}. Leptons of fourth
generation (E,N) should be very weakly mixed with the ordinary
ones, while in quark sector (U,D) mixing is limited only by
unitarity of $3\times 3$ CKM matrix. In particular it was noticed
in ref. \cite{4} that the predicted mass of the higgs could be as
high as 500 GeV. That conclusion was based on a sample of 10.000
random inputs of masses of fourth generation leptons and quarks.
However the sets of the lepton and quark masses were presented
independently (see Fig. 7 in ref. \cite{4}). Thus it is not clear
how they were combined.

In this letter we try to develop a systematic approach to the
problem by using our LEPTOP code \cite{33} to find steep and flat
directions in the five-dimensional parameter space: $m_H$, $m_U$,
$m_D$, $m_E$, $m_N$.
For each point in this space we perform three-parameter fit
($m_t, \alpha_s, \bar{\alpha}$) and calculate the $\chi^2$ of the fit.
 It turns out that the $\chi_{\rm min}^2$
depends weakly on $m_U +m_D$ and $m_H$, while its dependence on
$m_U - m_D$, $m_E$ and $m_N$ is strong. We limit ourselves to the
values of $m_N$ larger than 50 GeV because according to
experimental data from LEP II on the emission of initial state
bremsstrahlung photons, $m_N > 50$ GeV at 95\% c.l. \cite{44, 45}.

We analyzed Summer 2001 precision data (ref. \cite{1} which are
also given in the Table 1 in ref. \cite{3}). Figures 1-4 show
$\chi_{\rm min}^2$ (crosses) and constant $\chi^2$ lines
corresponding to $\Delta \chi^2 = 1, 4, 9, 16,$ ... on the plane
$m_N, m_U -m_D$ for fixed values of $m_U +m_D = 500$ GeV, $m_H =
120$ (Figs. 1 and 3) and 500 GeV (Figs. 2 and 4) and $ m_E=100$
(Figs. 1 and 2) and $300$ GeV (Figs 3 and 4). We also performed
fits for $m_H=300$ GeV.

The above choice of masses is based on a large number of fits
covering a broad space of parameters: 300 GeV $< m_U + m_D <$ 800
GeV; 0 GeV $< m_U - m_D <$ 400 GeV; 100 GeV $< m_E <$ 500 GeV; 50
GeV $< m_N <$ 500 GeV;  120 GeV $< m_H <$ 500 GeV. Concerning
quarks, $m_U + m_D$ is bounded from below by direct searches
limit, while from above by triviality arguments. Since $\chi^2$
dependence on $m_U + m_D$ is very weak, our choice of intermediate
value $m_U + m_D = 500$ GeV represents a typical, almost general
case. For this choice $|m_U - m_D|$ can not be larger than $\sim
200$ GeV because of the mentioned above direct searches bound.

Concerning charged lepton, its mass is taken above LEP II bound.
We present fits at two values of $m_E$ (100 GeV and 300 GeV) and
one can see how fit is worsening with $m_E$ going up.

Concerning the value of $m_H$, we vary it from the lower LEP II
limit up to triviality bound and since the dependence of
observables on $m_H$ is flat, one can get $\chi^2$ behaviour from
two limiting points: $m_H = $120 and 500 GeV.

For $m_E=100$ GeV we have the minimum of $\chi^2$ at
 $m_N \simeq 50$ GeV and:

\begin{center}
\begin{tabular}{lll}
for $m_H = 120$ GeV: & $|m_U -m_D| \sim 50$ GeV, &
$\chi_{\rm min}^2/n_{d.o.f.} = 20.6/12$  \\
for $m_H = 300$ GeV: & $|m_U -m_D| \sim 75$ GeV, &
$\chi_{\rm min}^2/n_{d.o.f.} = 20.8/12$ \\
for $m_H = 500$ GeV: & $|m_U -m_D| \sim 85$ GeV, &
$\chi_{\rm min}^2/n_{d.o.f.} = 21.4/12$
\end{tabular}
\end{center}

Thus we have two lines ($m_U>m_D$ and $m_U<m_D$) in the ($m_U -
m_D, m_H$) space that correspond to the best fit of data. Along
these lines the quality of the fit is only slightly better for
light higgs ($m_H \sim 120$ GeV) than for the heavy one ($m_H
\sim$ 300 -- 500 GeV).

Note that the $n_{d.o.f}$ is 12, unlike the case of the Standard
Model, where it was 13 (ref. \cite{3}). This change occurs because
in the present paper $m_H$ is not a fitted, but a fixed parameter
(hence 13 becomes 14), while $m_N$ and $m_U - m_D$ are two
additional fitted parameters (hence 14 becomes 12). (As is well
known, $n_{d.o.f.}$ is equal to the number of experimentally
measured observables minus the number of fitted parameters.)

For $m_E=300$ GeV we have the minimum of $\chi^2$ at
 $m_U-m_D \simeq 25$ GeV and:

\begin{center}
\begin{tabular}{lll}
for $m_H = 120$ GeV: & $m_N \sim 200$ GeV, &
$\chi_{\rm min}^2/n_{d.o.f.} = 23.0/12$  \\
for $m_H = 300$ GeV: & $m_N \sim 170$ GeV, &
$\chi_{\rm min}^2/n_{d.o.f.} = 24.0/12$ \\
for $m_H = 500$ GeV: & $m_N \sim 150$ GeV, &
$\chi_{\rm min}^2/n_{d.o.f.} = 24.4/12$
\end{tabular}
\end{center}

Thus, the best fit of the data corresponds to the light $m_E
\simeq 100$ GeV and $m_N\simeq 50$ GeV. The significance of light
$m_N$ (around 50 GeV) was first stressed in \cite{111}. Increase
of $m_E$ leads to the increase of $m_N$ and to fast worsening of
$\chi_{\rm min}^2$.

Although inclusion of one extra generation improves the
quality of the fit (compare $\chi^2/n_{d.o.f.} = 23.8/13$ for the
SM from \cite{3} and $\chi_{\rm min}^2/n_{d.o.f.} = 20.6/12$ from Fig. 1) it
remains pretty poor. The poor quality of the fit is due to $3.3
\sigma$ discrepancy in $s_l^2 \equiv\sin^2\theta_{\rm eff}$
extracted from leptonic decays and from $A_{FB}^{b,c}$ \cite{5}. If one
multiplies experimental errors of $A_{FB}^b$ and $A_{FB}^c$ by a
factor 10, one gets good quality of SM fit \cite{5, 3} but with
extremely light
higgs, having only a small (few percent) likelihood to be consistent with
the lower limit from direct searches.
 We prove that the fourth generation allows to have
higgs as heavy as 500 GeV with a perfect quality of the fit:
$\chi_{\rm min}^2/n_{d.o.f.} = 13/12$, if one uses old NuTeV data
(see caption of Fig. 2).

To qualitatively understand the dependence of $m_U -m_D$ on
$m_H$ in the case of $m_E = 100$ GeV at $\chi_{\rm min}^2$ let us
recall how radiative
corrections to the ratio $m_W/m_Z$ and to $g_A$ and $R=g_V/g_A$ (the axial and the
ratio of vector and axial couplings of $Z$-boson to charged
leptons) depend on these quantities \cite{9}:
\begin{equation}
\delta V^i\approx\left[-\left(
\begin{array}{c}
\frac{11}{9}s^2 \\ s^2 \\ s^2 +\frac{1}{9}
\end{array}
\right)\ln(\frac{m_H}{m_Z})^2 +\frac{4}{3}\frac{(m_U
-m_D)^2}{m_Z^2}+\left(
\begin{array}{c}
\frac{16}{9}s^2\frac{m_U -m_D}{m_U + m_D} \\ 0 \\ \frac{2}{9}
\frac{m_U -m_D}{m_U + m_D}
\end{array}
\right)\right] \label{1}
\end{equation}
where $i=m, A, R$, while $s^2 \simeq 0.23$. Corrections to other
observables can be calculated in terms of $\delta V^i$. In the
vicinity of $\chi_{\rm min}^2$ the third term in brackets is much
smaller than the second one. Hence the smallness of the left-right
asymmetry of the plots of Figs. 1, 2.  Since $\frac{11}{9}s^2
\approx s^2 +\frac{1}{9} \approx s^2$, the increase of $m_H$ is
compensated by increase of $|m_U -m_D|$ and we have a valley of
$\chi_{\rm min}^2$.

Captions of Figs. 1 and 2 reflect recent change in NuTeV data
(from $m_W = 80.26 \pm 0.11$ GeV
\cite{10} to $m_W = 80.14 \pm 0.08$ GeV \cite{11}) which results
in drastic worsening of the fit even in the presence of the
fourth generation.

Thus we see that the 4th family scenario is better than the
Standard Model, because the latter can produce good fit only when
the mass of the higgs is much lower than the lower limit of LEP
II, even when experimental data on heavy quark asymmetries and new
NuTeV data are ignored.

Note that originally introduced in \cite{1111} parameters $S, T,
U$ are not adequate for the above analysis, because they assume
that all particles of the fourth generation are much heavier than
$m_Z$, while in our case the best fit corresponds to $m_N \sim
m_Z/2$. In the paper \cite{4} modified definitions of $S$ and $U$
were used in order to deal with new particles with masses
comparable to $m_Z$. However, let us stress that both original and
modified definitions of $S$, $T$ and $U$ take into account
radiative corrections from the ``light'' 4th neutrino only
approximately, while the threshold effects, that are so important
for $m_N \simeq 50$ GeV, can be adequately described in the
framework of functions $V^i$.

In conclusion let us stress that in the framework of SUSY with
three generations radiative corrections due to loops with
superpartners also shift upward the mass of the higgs in the case
of not too heavy squarks (300-400 GeV, see Table 1
 in  \cite{gaidaenko1999}) or light
sneutrinos (55-80 GeV, see  \cite{altarelli2001}).

V.N., L.O. and M.V. were partly supported by RFBR grant No.
00-15-96562; V.N. was partly supported by the grant INTAS OPEN
2000-110 as well. We are grateful to M.Chanowitz for his comments.

\begin{figure*}[]
\centering
\includegraphics[width=0.84\textwidth]{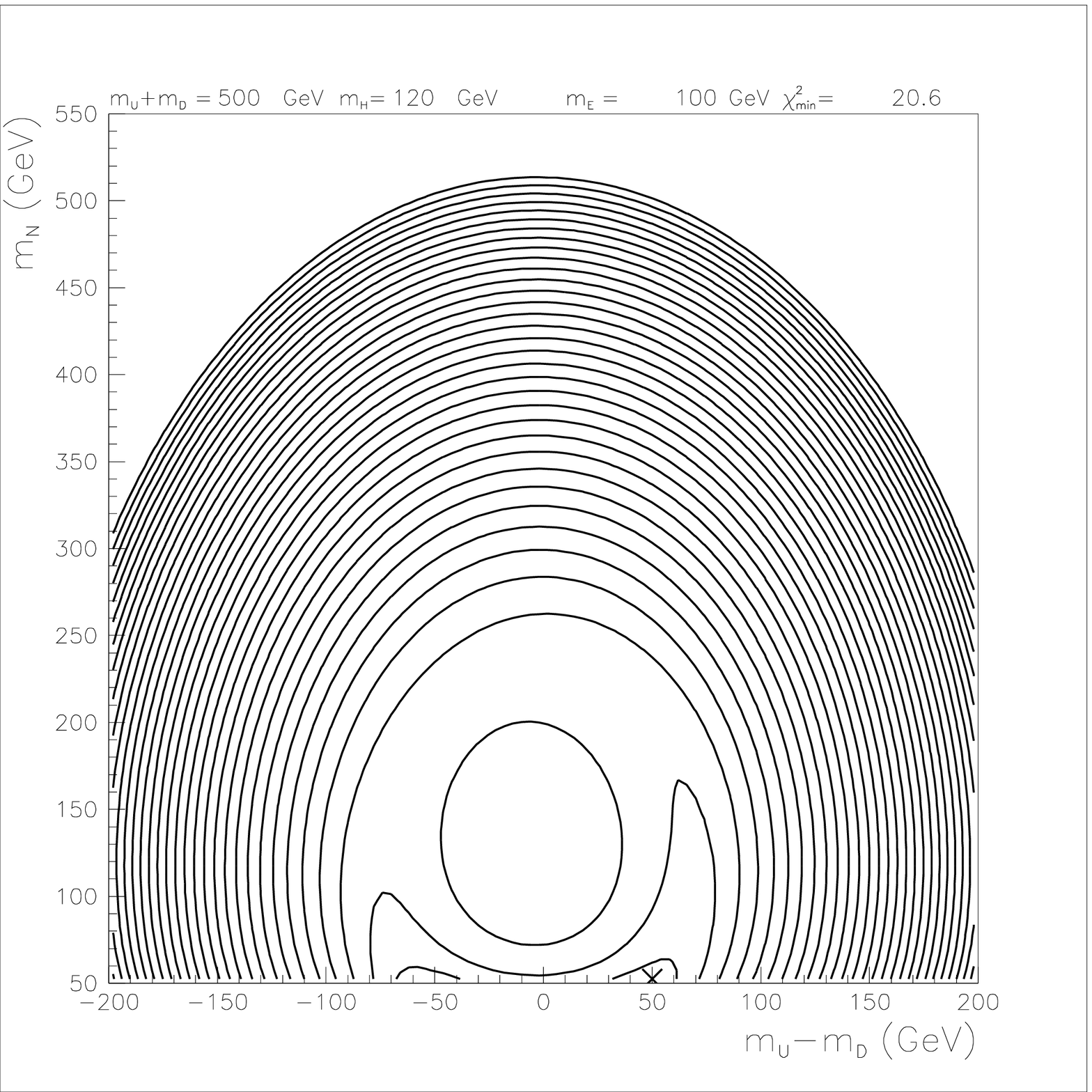}
\caption{\label{FIG1} Exclusion plot
on the plane $m_N, m_U -m_D$ for fixed values of $m_H=120$ GeV,
$m_U +m_D = 500$ GeV and $m_E=100$ GeV.
 $\chi_{\rm min}^2$ shown by two crosses corresponds to $\chi^2/n_{d.o.f.} =
20.6/12$. (The left-hand cross is slightly below $m_N = 50$ GeV.)
Borders of regions show
domains allowed  at the level $\Delta \chi^2 = 1, 4, 9, 16$, etc.
 The plot was based on the old NuTeV data. The new NuTeV data preserve
the pattern of the plot, but lead to
$\chi_{\rm min}^2/n_{d.o.f.} = 27.7/12$. If  $A_{FB}^b$ and $A_{FB}^c$
uncertainties are
multiplied by factor
$10$ we get $\chi_{\rm min}^2/n_{d.o.f.} = 19.1/12$ for  new NuTeV, and
$\chi_{\rm min}^2/n_{d.o.f.} = 11.3/12$ for old NuTeV with practically
 the same pattern of the plot.}
\end{figure*}

\begin{figure*}[]
\centering
\includegraphics[width=0.84\textwidth]{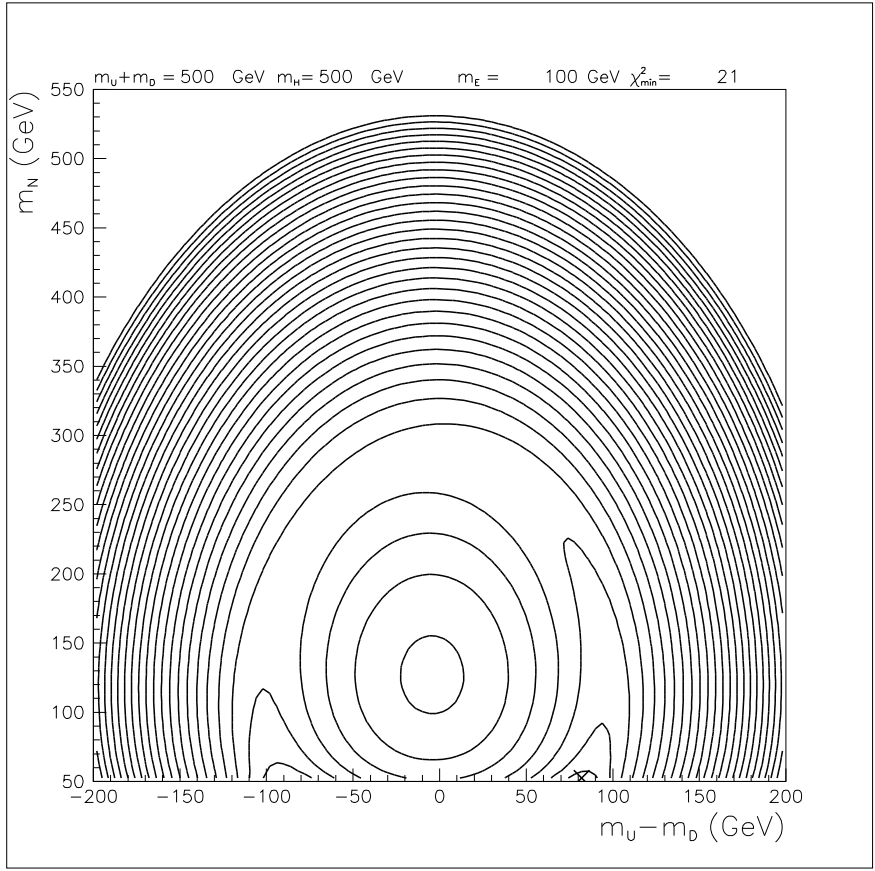}
\caption{\label{FIG3} Exclusion plot
on the plane $m_N, m_U -m_D$ for fixed values of $m_H=500$ GeV,
$m_U +m_D = 500$ GeV and $m_E=100$ GeV.
 $\chi_{\rm min}^2$ shown by two crosses corresponds to $\chi^2/n_{d.o.f.} =
21.4/12$. (The left-hand cross is slightly below $m_N = 50$ GeV.)
Borders of regions show
domains allowed  at the level $\Delta \chi^2 = 1, 4, 9, 16$, etc.
The plot was based on the old NuTeV data. The new NuTeV data preserve
the pattern of the plot, but lead to
$\chi^2_{\rm min}/n_{d.o.f.} = 28.3/12$. If  $A_{FB}^b$ and $A_{FB}^c$
uncertainties are
multiplied by a factor
$10$, we get $\chi^2_{\rm min}/n_{d.o.f.} = 21.2/12$ for  new NuTeV, and
$\chi^2_{\rm min}/n_{d.o.f.} = 13/12$ for
old NuTeV with practically the same pattern of the plot.}
\end{figure*}

\begin{figure*}[]
\centering
\includegraphics[width=0.84\textwidth]{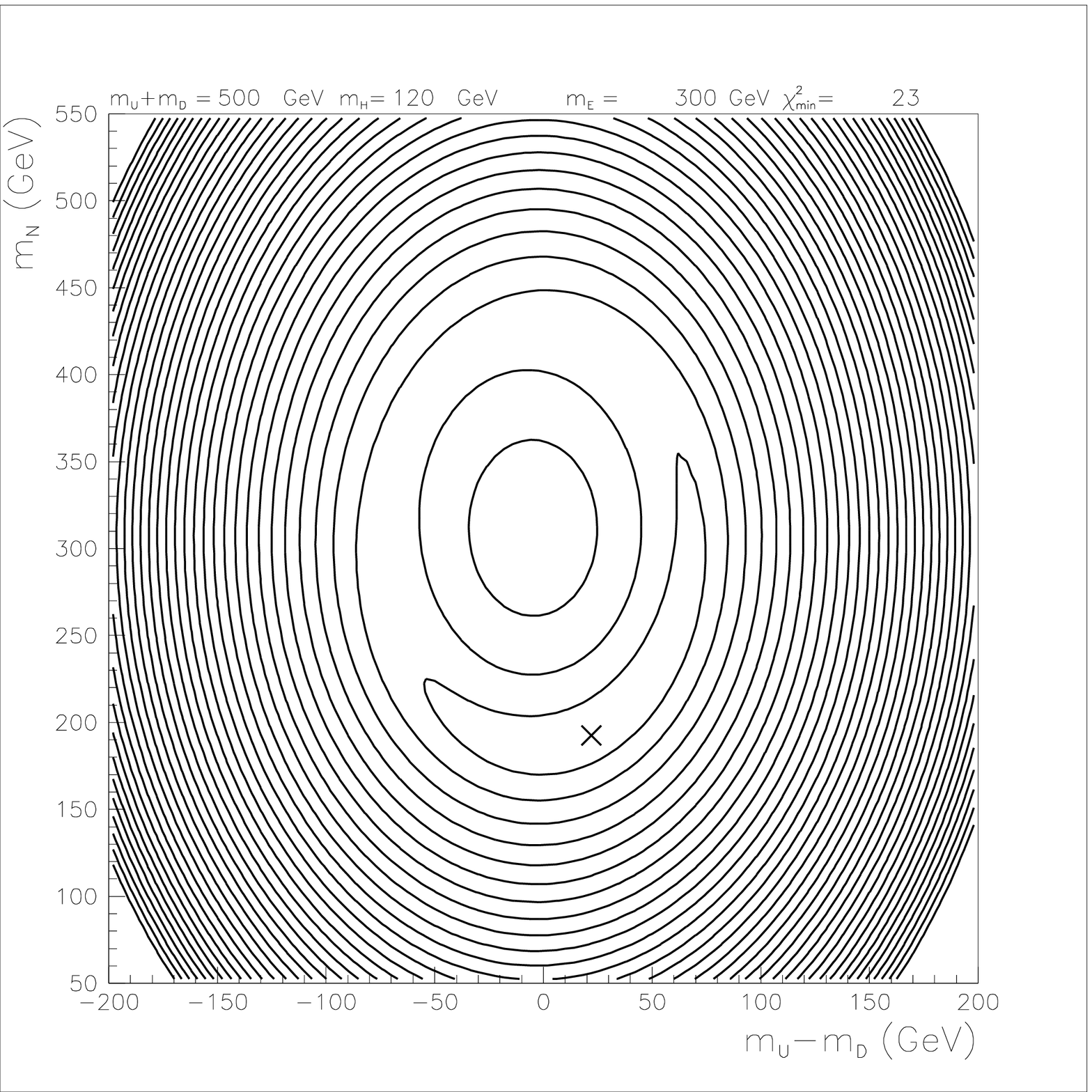}
\caption{\label{FIG4} Exclusion plot
on the plane $m_N, m_U -m_D$ for fixed values of $m_H=120$ GeV,
$m_U +m_D = 500$ GeV and $m_E=300$ GeV.
 $\chi_{\rm min}^2$ shown by cross corresponds to $\chi^2/n_{d.o.f.} =
23.0/12$. Borders of regions show
domains allowed  at the level $\Delta \chi^2 = 1, 4, 9, 16$, etc. }
\end{figure*}

\begin{figure*}[]
\centering
\includegraphics[width=0.84\textwidth]{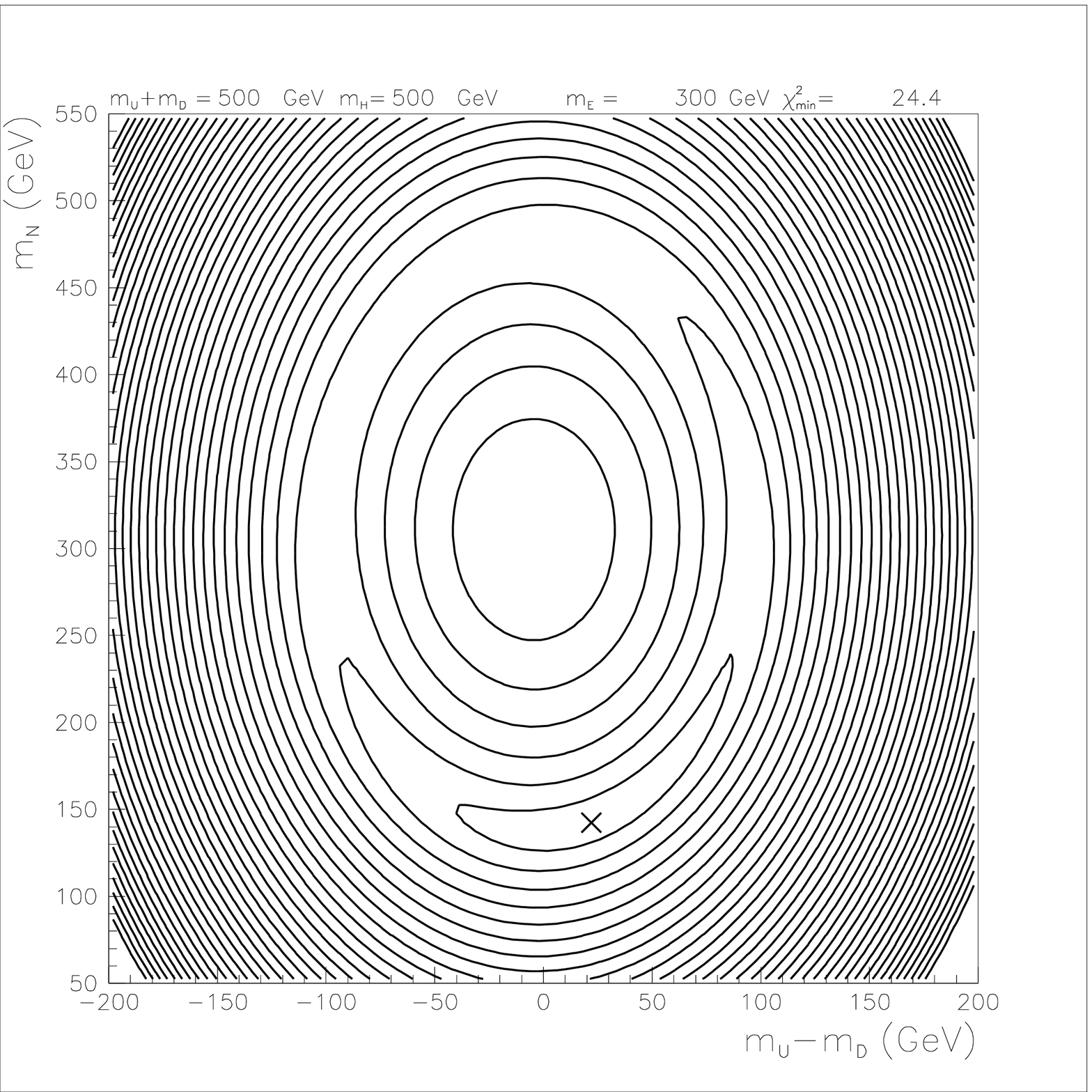}
\caption{\label{FIG6} Exclusion plot
on the plane $m_N, m_U -m_D$ for fixed values of $m_H=500$ GeV,
$m_U +m_D = 500$ GeV and $m_E=300$ GeV.
 $\chi_{\rm min}^2$ shown by cross corresponds to $\chi^2/n_{d.o.f.} =
24.4/12$. Borders of regions show
domains allowed  at the level $\Delta \chi^2 = 1, 4, 9, 16$, etc. }
\end{figure*}

\end{document}